\documentstyle[twocolumn,prl,aps,epsf]{revtex}
\begin{document}
\draft
\newcommand{\disc}{\mbox{\scriptsize disc}}
\newcommand{\cont}{\mbox{\scriptsize cont}}

\title{Connection between energy--spectrum self--similarity \\
       and specific heat log--periodicity}

\author{ Ra\'ul O. Vallejos$^{1,2}$, 
         Renio S. Mendes$^{2,3}$, 
         Luciano R. da Silva$^{2,4}$, 
         and \\
         Constantino Tsallis$^{2}$ }

\address{
$^1$Instituto de F\'{\i}sica, UERJ, R. S\~ao Francisco Xavier 524, 
20559-900-Rio de Janeiro-RJ, Brazil. \\
$^2$Centro Brasileiro de Pesquisas F\'\i sicas, R. Xavier Sigaud 150,
22290-180 Rio de Janeiro-RJ, Brazil.\\
$^3$Departamento de F\'\i sica, Universidade Estadual de Maring\'a, 
Av. Colombo 5790, 87020-900 Maring\'a-PR, Brazil. \\
$^4$Departamento de F\'\i sica, Universidade Federal do Rio Grande do 
Norte, 59072-970 Natal-RN, Brazil.}

\date{\today}
\maketitle

\begin{abstract} 
As a first step towards the understanding of the thermodynamical
properties of quasiperiodic structures, we have performed both
analytical and numerical calculations of the specific heats associated
with successive hierarchical approximations to multiscale fractal
energy spectra. We show that, in a certain range of temperatures, the
specific heat displays log--periodic oscillations as a function of the
temperature. We exhibit scaling arguments that allow for
relating the mean value as well as the amplitude and the period of the
oscillations to the characteristic scales of the spectrum.
\end{abstract}

\section{Introduction}

The discovery of quasicrystals in 1984\cite{discovery} aroused a great
interest in quasiperiodic structures, as is confirmed by the great
number of theoretical\cite{theor} and experimental\cite{exp} works that
followed (see also \cite{books}).  In particular, the behavior of a
variety of particles and quasi--particles (electrons \cite{electrons},
phonons\cite{phonons}, and others\cite{truchons}) in quasiperiodic
structures has been and is currently being studied.  A {\it fractal
energy spectrum} is a common feature to such structures (e.g.,
\cite{fibonacci}). As in general these spectra tend to be quite
complex, simple models have been studied to enlighten the
thermodynamical specificities that such systems may display.  Within
this vein, we studied in \cite{tsallis} one of the simplest fractal
spectra (the triadic Cantor set); there it was shown that the specific
heat of such system exhibits a very particular behaviour: it oscillates
log--periodically around a mean value which equals the fractal
dimension of the spectrum.
 
In this paper we extend the analysis of \cite{tsallis} to the case of
multi--scale spectra and present a theoretical connection between the
structure of the spectrum and the corresponding thermodynamical
properties.  In particular, we make explicit the relationship between
discrete scale--invariance and log--periodic oscillations.  We start
our discussion with the two scale $(r_1,r_2)$ fractal set, in its
discrete version (Section II).  Then, in Section III, we show that the
results of II can be extended to the continuous two--scale case, as
well as to the $n$--scale problem.  Section IV contains the
conclusions.

\section{Two scale discrete model}

Let us begin by considering a spectrum lying on the $(r_1,r_2)$ two
scale fractal set \cite{Tel88} indicated in Fig.~1a (the restrictions
$0 \le r_1,r_2,r_1+r_2 \le 1$ apply).  The starting point ($n=0$) for
the construction of this spectrum is an arbitrary discrete set of
levels lying in an energy interval we take to be $[0,1]$. This is the
pattern that will be repeated ad infinitum in a self--similar way. 
\begin{figure}[ht]
\hspace{0.2cm}
\epsfxsize=7.5cm
\epsfbox[91 255 516 471]{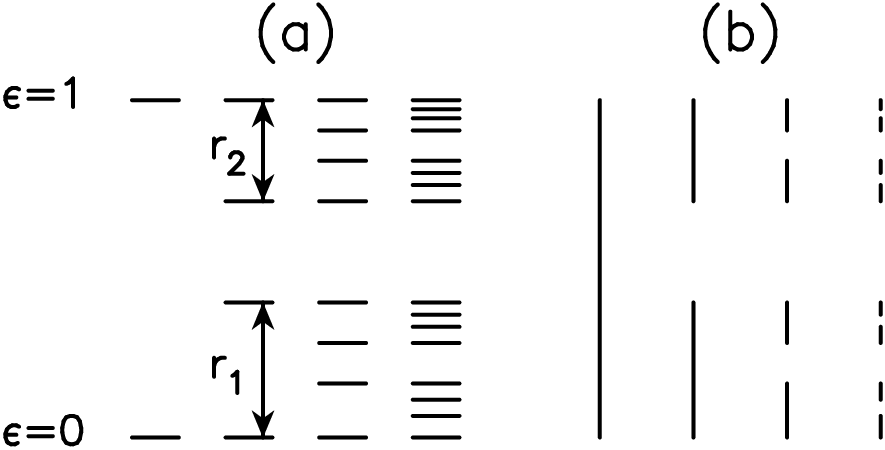}
\vspace{0.5cm}
\caption{Energy spectra. 
The first four steps in the construction of $(r_1,r_2)$ fractal sets.
We show a discrete case (a) in which the starting point ($n=0$) is a
set of two levels at $\epsilon=0,1$.  These levels are then compressed
by a factor $r_1$ ($r_2$) and put on bottom (top) of the interval
$[0,1]$ ($n=1$), and so on for increasing $n$. The construction of a
continuous example (b) starts from a band of uniform density in
$[0,1]$. The iterative rule is the same as in (a), i.e.  $n=1$
corresponds to a spectrum whose first and second bands are the
intervals $[0,r_1]$ and $[1-r_2,1]$ respectively, etc. We take the
level density inside each band to be a constant, and the same for all
bands in a given hierarchy.  In both cases (a) and (b), a fractal
emerges at the $n \rightarrow \infty$ limit.}
\end{figure}
The next step consists in compressing the set $n=0$ by factors $r_1$
and $r_2$ and putting the two resulting pieces on bottom and top of the
interval $[0,1]$, respectively (see Fig.~1a).  Recursive application of
this rule eventually leads to a set of fractal dimension $d_f$ given by
$r_1^{d_f}+r_2^{d_f}=1$ (hence, if $r_1=r_2 \equiv r$, $d_f=-\ln 2/\ln
r$).  This rule can be explicitly written as a recurrence equation for
the energy levels $\epsilon_j^{(n)}$ at the $n$--th stage,
\begin{equation}
\left\{ \epsilon_j^{(n+1)} \right\} = 
                \left\{r_1\epsilon_j^{(n)}\right\} \cup
  \left\{ 1-r_2 + r_2\epsilon_j^{(n)}\right\} ~.
\label{levels}
\end{equation}
This analytical rule for the construction of the spectrum is the key
to obtaining scaling relations for the thermodynamical quantities.
The starting point is the partition function for a given hierarchy 
$n$: 
\begin{equation}
Z^{(n)}(\beta) = \frac{1}{2^n} \sum_{j=1}^{2^n} 
          \exp(-\beta \epsilon_j^{(n)}) ~,
\label{partition}
\end{equation}
where $\beta$ is the inverse temperature (we are considering a unit
Boltzmann constant, i.e., $k_B=1$).  A normalization prefactor has been
included so that $Z^{(n)}$ is well defined in the limit $n \rightarrow
\infty$; in any case, it will not affect the thermodynamics of the
system. Now, a recurrence formula for the partition function is readily
obtained as a direct consequence of the self--similarity of the energy
set (\ref{levels}):
\begin{equation}
Z^{(n+1)}(\beta) = \left[ Z^{(n)}(\beta r_1) +
                          e^{-\beta (1-r_2)} Z^{(n)}(\beta r_2)    
                  \right]/2 ~.
\end{equation}
Introducing 
$Z(\beta) \equiv \lim_{n \rightarrow \infty} Z^{(n)}(\beta)$, 
we have 
\begin{equation}
Z(\beta) = \left[ Z(\beta r_1) + e^{-\beta (1-r_2)} 
                  Z(\beta r_2)  \right]/2 ~.
\label{recZ}
\end{equation}
From now on we will restrict our discussion to the low temperature
regime, as this is the most interesting one. In fact, as the
temperature is lowered, the smaller scales of the fractal are
progressively revealed, and anomalous effects are expected.  Moreover,
in the case $T \ll 1$ the analysis gets simplified and some general
conclusions can be obtained.  In this regime we can safely neglect the
exponentially small term in (\ref{recZ}) and derive scaling relations
for the partition function, the dimensionless free energy $Q \equiv
F/T=-\ln Z$, the total energy $E$, the entropy $S$, and the specific
heat $C$:
\begin{eqnarray}
Z(T) & = & Z(T/r_1)/2         \\
Q(T) & = & Q(T/r_1) + \ln 2   \\
E(T) & = & r_1 E(T/r_1)       \\
S(T) & = & S(T/r_1) - \ln 2   \\
C(T) & = & C(T/r_1)           ~.
\end{eqnarray}
{\em Independently of the $n=0$ energy pattern}, the relevant scale
factor is $r_1$ (conversely, $r_2$ governs the scaling laws for {\em
negative} temperatures).
 
The most interesting of the equalities above is the last one, which
expresses the fact that the specific heat is a 
log--periodic function
of the temperature, that is $C(T)=f(2 \pi \ln T/\ln r_1)$, where $f$ is
a $2\pi$-periodic function.  In other words, if one sets $T=r_1^x$, $C(x)$
results in a periodic function of $x$ (of period one).  Consistently,
its mean value can be calculated as
\begin{eqnarray}
\langle C(T) \rangle  & = & \int_{x_0}^{x_0+1} C(r_1^x) \, dx = 
    -\frac{1}{\ln r_1}\int_{r_1\tau}^{\tau} C(T) \, 
     \frac{dT}{T} \nonumber \\ 
                      & = &
    -\frac{S(\tau)-S(r_1\tau)}{\ln r_1} = -\frac{\ln 2}{\ln r_1} ~.
\label{meanheat}
\end{eqnarray}
In Ref.~1 it was shown that for a one--scale Cantor spectrum (i.e.
$r_1=r_2=1/3$), the average value $\langle C(T) \rangle$ coincides with
the fractal dimension $d_f$.  The equality above shows that {\em this
is not the case for a two scale fractal}. We will come back to this
point later on to argue that the ``dimension'' $d=-\ln 2/\ln r_1$ can
be given a simple meaning.  We remark that the result (\ref{meanheat})
holds as long as $T \ll 1$ (we recall that, as the spectrum is
bounded, for high temperatures the specific heat must decay as
$T^{-2}$).

The scaling reasoning has given us information concerning the mean
values. In order to discuss the oscillations around the mean value, we
resort to a numerical analysis. Starting from (\ref{partition}) we
have computed finite approximations to $C(T)$,
\begin{equation}
C^{(n)}(T)=\frac{\partial}{\partial T} \left[
           T^2 \frac{\partial \ln Z^{(n)} }{\partial T} \right]  ~,
\end{equation}
and studied its dependences on the hierarchical depth $n$, and the
parameters $r_1$ and $r_2$. In Fig.~2 we show some plots of the
specific heat vs. temperature for different values of $(r_1,r_2)$ and
fixed hierarchical depth $n=8$.
\begin{figure}[ht]
\hspace{0.0pc} 
\epsfxsize=7.5cm
\epsfbox[95 114 483 594]{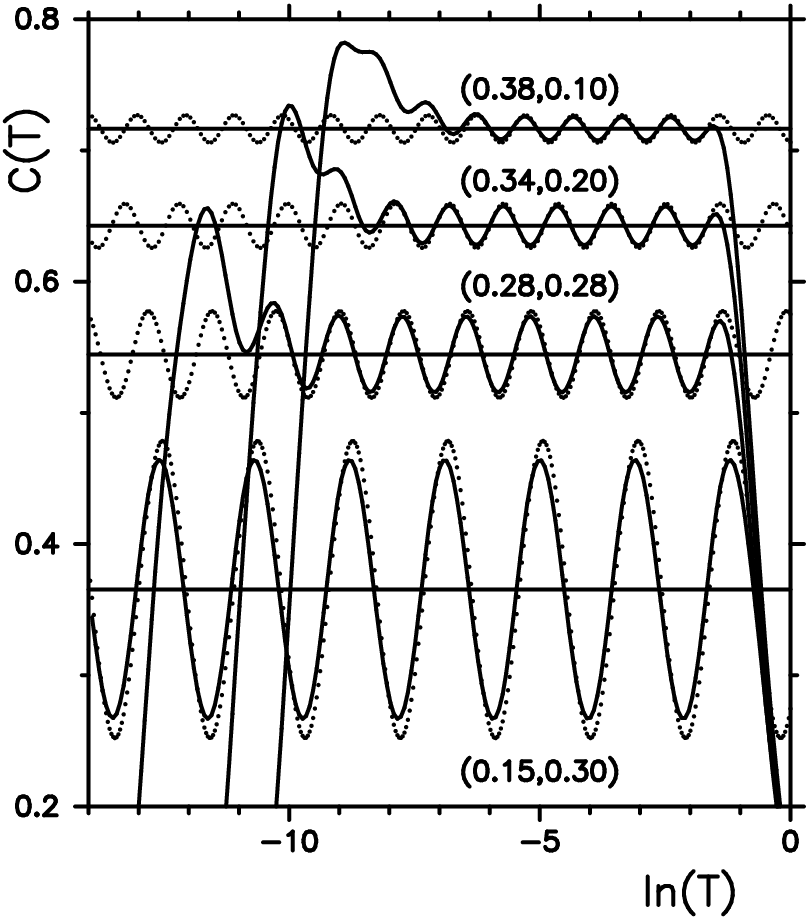}
\vspace{-1.0pc} 
\caption{Specific heat (in units of $k_B$) vs. temperature 
(in units of the width of the spectrum) for the $(r_1,r_2)$ discrete
fractal set of Fig.~1 ($n=8$).  Two levels at $\epsilon=0,1$ were taken
as the $n=0$ pattern.  The curves are parametrized by the scale factors
$(r_1,r_2)$.  The horizontal lines indicate the average value $\langle
C \rangle = d = -\ln 2 / \ln r_1$. The dotted lines correspond to our
prediction $C \approx d + a'' \cos(\omega \ln T) + b'' \sin(\omega \ln
T)$, where $\omega=-2\pi/\ln r_1$.  The parameters $a''$, $b''$ are
related to basic properties of the smoothed spectrum (see text).  For
high temperatures ($\ln T >0$) the specific heat decays as $T^{-2}$,
for arbitrary $n$.  The low--temperature breakdown of the oscillatory
behaviour is pushed towards the left when $n$ increases.}
\end{figure}
It is apparent that there is range of
temperatures in which $C(T)$ behaves in the way the above scaling
arguments predict.  This is a range of intermediate temperatures,
$T_{\min} \ll T \ll 1$ where $T_{\min} \sim r_1^{n} $ is
associated to the smallest scale of the ``truncated fractal''; of
course, $T_{\min}\rightarrow 0$ in the limit $n \rightarrow \infty$.
Fig.~2 clearly displays the following features. $C(T)$ oscillates
log--periodically around the mean value $d=-\ln 2/\ln r_1$ with
frequency $\omega=-2\pi/\ln r_1$. Notice that each curve completes
about $n$ periods ($n=8$ in the figure but we have verified this
behaviour for higher depth $n$ as well).  
In Fig.~3a we show a typical
example illustrating the dependence of the amplitudes of the specific
heat on the scales $(r_1,r_2)$. We have plotted the maximum and minimum
values of the low temperature regime of $C(T)$ (denoted respectively
$C^+$ and $C^-$), together with the mean value $\langle C \rangle =-\ln
2/\ln r_1$ for the family $r_1+r_2=2/3$.  The oscillations decrease in
amplitude as $r_1$ decreases, and for $r_1$ sufficiently small ($r_1
\stackrel{<}{\sim} 0.1$) assymetries become significant.

A point of view that allows for understanding quantitatively the
amplitudes of the oscillations consists in relating the thermodynamical
properties to those of the spectrum. For instance, a constant value of
the specific heat $C=\sigma$ is associated in general to the fact that
the cummulative density of states (or spectral staircase)
\begin{equation}
N(\epsilon)=\int_0^\epsilon \rho(x) dx
\end{equation}
scales with energy as $\epsilon^\sigma$ (equipartition principle). In
our case it can be verified that the spectral staircase grows
approximately as $\epsilon^d$ (see Fig.~4a), and consequently the
average specific heat is $\langle C \rangle = d$. It is also apparent
from Fig.~4b that the integrated density of states $N(\epsilon)$ is a
log--periodic function of the energy.  (Similar results were obtained
by Kimball and Frisch for the distribution of normal mode frequencies
of fractal--based models \cite{kimball}; see also \cite{petri}). In
fact, this features could have been anticipated by noting that
$N(\epsilon)$ also satisfies a simple scaling law:
\begin{equation}
N(r_1\epsilon)=N(\epsilon)/2,
\end{equation}
whose general solution can be written as a power law times a 
log--periodic function\cite{erzan}.

We will now show that an excellent description of the specific heat can
be achieved by considering the first non--trivial correction to the
bare power-law scaling for $N(\epsilon)$:
\begin{equation}
N(\epsilon) \approx \epsilon^{d}
                  [a+b\cos(\omega \ln \epsilon - \phi)] ~.
\label{smoothN}
\end{equation}
In principle there is not a unique criterion for choosing the
parameters $a$, $b$, $\phi$.  We have determined $a$ and $b$ by
requiring that the exact $N(\epsilon)$ and its smooth approximation
(\ref{smoothN}) have the same average value and variance. The condition
of maximum overlap between the exact staircase fluctuations
$N/\epsilon^d$ and the cosine function fixes $\phi$.
\begin{figure}[h]
\epsfxsize=7.5cm
\epsfbox[101 155 486 594]{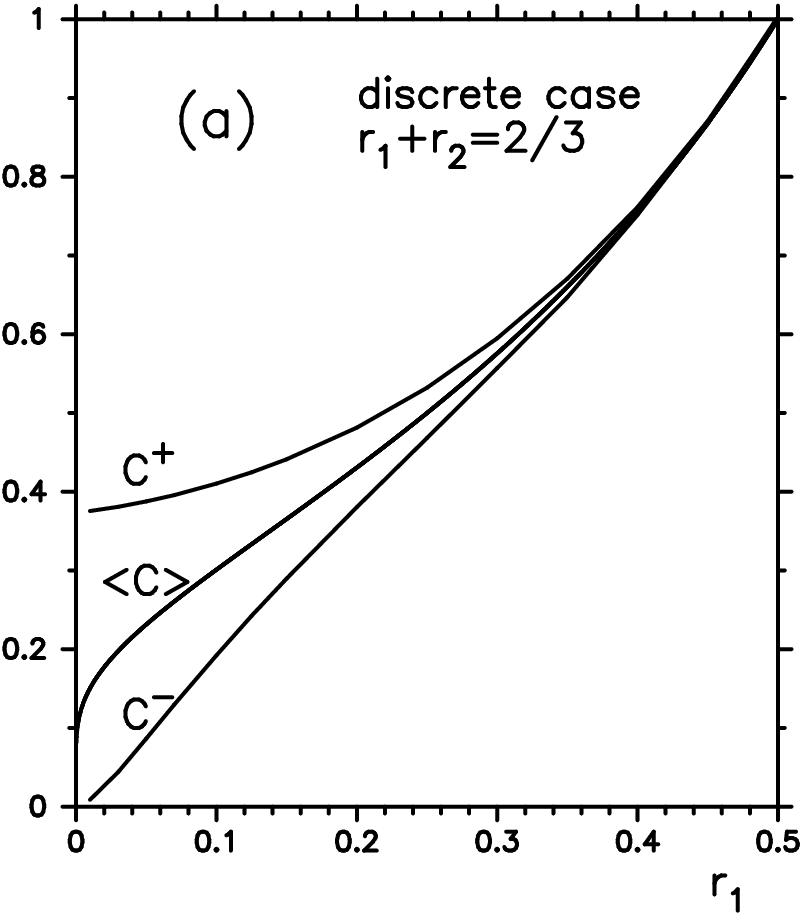}
\epsfxsize=7.5cm
\epsfbox[114 114 486 594]{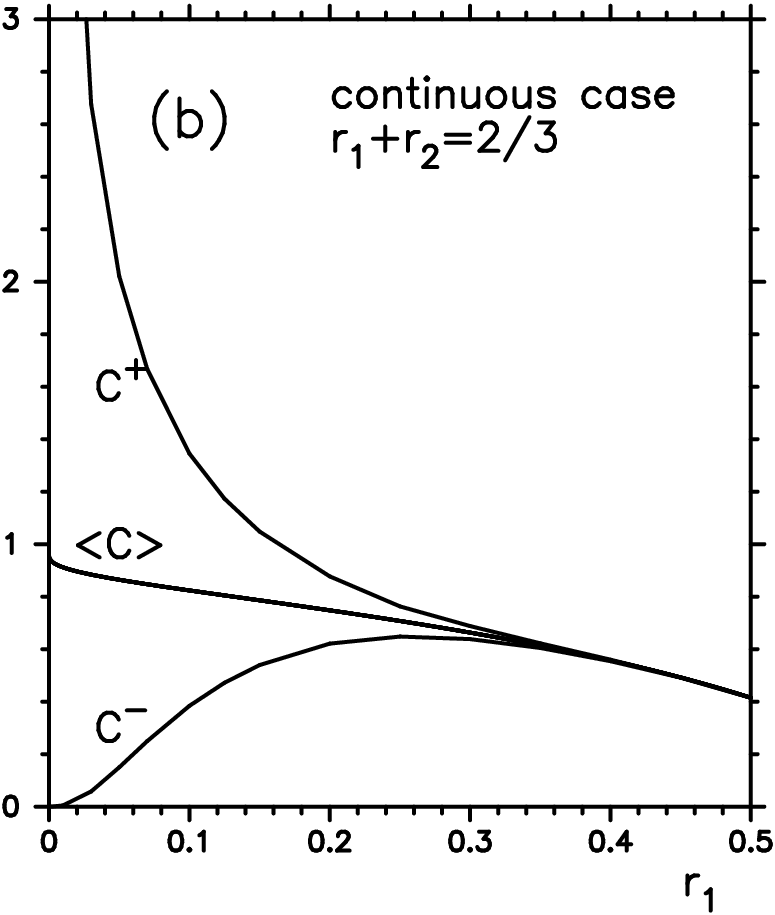}
\caption{Amplitudes of the oscillations of the specific heat in
the log-periodic regime ($T \ll 1$). 
Maximum and minimum values of $C(T)$ (respectively $C^+$ and $C^-$) 
for the family of fractal spectra $r_1+r_2=2/3$. 
Also shown is the average specific heat 
$\langle C \rangle$. For the discrete case (a), 
$\langle C \rangle =-\ln 2/\ln r_1$. In the continuous case (b),
$\langle C \rangle =1-\ln (r_1+r_2)/\ln r_1$.}
\end{figure}
An approximate
partition function is now written in terms of the smoothed cummulative
level density,
\begin{eqnarray}
Z(T) & \approx & \beta \int_0^\infty 
                 \epsilon^{d}[a+b\cos(\omega \ln \epsilon - \phi)] 
                 \exp(-\beta \epsilon) d\epsilon   \\ \nonumber
     &    =    & T^d \, [ aa' + 
                          bb' \cos(\omega \ln T - \phi) - 
                          bc' \sin(\omega \ln T - \phi)   ]  ~.
\end{eqnarray}
Here the constants $a',b',c'$ are calculated as the integrals:
\begin{equation}
\left\{ \matrix{ a' \cr 
                 b' \cr 
                 c' \cr} \right\} \equiv
\int_0^{\infty} dx x^d e^{-x} 
\left\{ \matrix{ 1                  \cr 
                \cos(\omega \ln x) \cr
                \sin(\omega \ln x) \cr} \right\}  ~.
\end{equation}
For not very small values of $r_1$ one has 
$a' \gg b',c'$ (e.g. if $r_1=0.34$, then
$a'\approx 0.9$, 
$b'\approx 0.002$, and 
$c'\approx 0.0003$.) 
After some straightforward manipulations,
to first order in the small parameters $b'/a'$ and $c'/a'$,
we obtain the specific heat
\begin{equation}
 C(T)  \approx    d 
                + a'' \cos(\omega \ln T)
                + b'' \sin(\omega \ln T)   ~.
\label{Csmooth}
\end{equation}
This expression can be seen as a log--Fourier expansion of the specific
heat up to second order terms.  Instead of presenting (complicated)
expressions for the constants $a''$ and $b''$ as functions of $r_1$ and
$r_2$ we prefer to show the specific heat (\ref{Csmooth}) for a set of
selected values of $r_1$, $r_2$ (Fig.~2, dotted lines). The agreement
of our approximation (\ref{Csmooth}) with the exact (numerical)
calculations is excellent for the three upper curves and reasonably
good for that corresponding to the smallest $r_1$ (higher order terms
might be necessary in this case).

\section{Continuous and multi-scale extensions}

For the sake of completeness let us also discuss the case of a spectrum
constructed by iterative use of the rule $(r_1,r_2)$ but starting from
a {\em continuous} pattern as shown in Fig.~1b.  For instance, if the
zero--th hierarchy is chosen as a continuous spectrum with uniform
density in the interval $[0,1]$, then $n=1$ corresponds to a spectrum
whose first and second bands are the intervals $[0,r_1]$ and
$[1-r_2,1]$ respectively; and so on for increasing $n$. We have chosen
the density to be uniform inside each band. In other words the number
of states in each band is proportional to its length, whereas in the
{\em discrete} case each ``band" contains the same number of states;
this is the essential difference between what we call {\em continuous}
and {\em discrete} spectra.  Now the partition function is written as
\begin{equation}
Z_{\cont}^{(n)}(\beta) =
\frac{1}{(r_1+r_2)^n} \int_0^1 \rho(\epsilon) \exp(-\beta \epsilon) 
d\epsilon 
\end{equation}
where, as in (\ref{partition}), a normalization prefactor has been
included.  The analogous to (\ref{levels}) is a recurrence equation for
the density of states:
\begin{equation}
\rho^{(n+1)}(\epsilon)=
\left\{ \matrix{ 
\rho^{(n)}(\epsilon/r_1)       & 
\mbox{if ~ $0 \le \epsilon \le r_1$  }\cr
           0                   & 
\mbox{if ~ $r_1 < \epsilon < 1 - r_2$}\cr     
\rho^{(n)}([\epsilon-1]/r_2+1) & 
\mbox{if ~ $1-r_2 \le \epsilon \le 1$}\cr} 
         \right. ~,
\label{Eqtrans}
\end{equation}
leading to the following result for the partition function 
\begin{equation}
Z_{\cont}(\beta) =  \frac{1}{r_1+r_2} \left[
                       r_1 Z_{\cont}(\beta r_1) + 
                       r_2 e^{-\beta (1-r_2)} Z_{\cont}(\beta r_2)
                                        \right]  ~.    
\end{equation}
\begin{figure}[ht]
\epsfysize=7.5cm
\epsfbox[73 172 483 573]{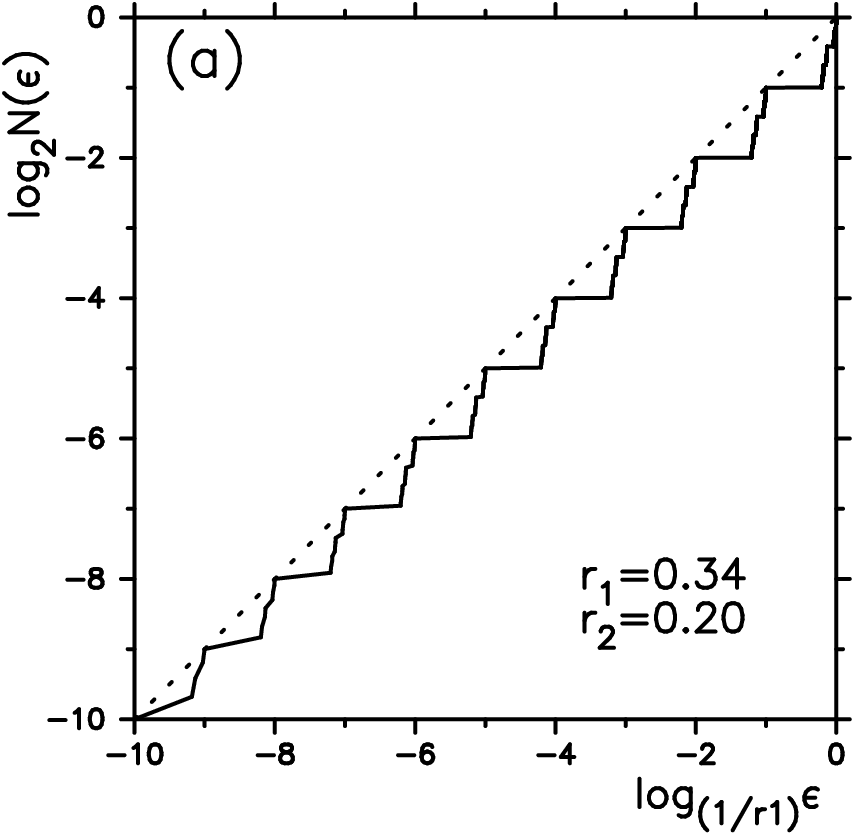}
\hspace{-0.5pc}
\epsfysize=8.2cm
\epsfbox[83 151 483 594]{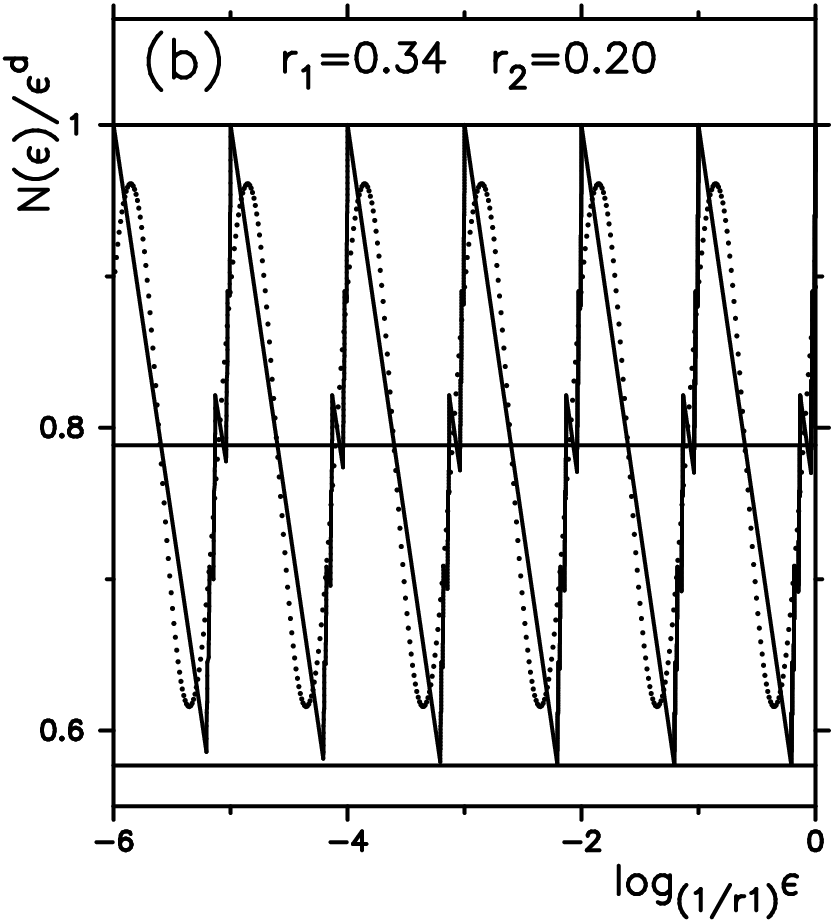}
\vspace{1.0pc}
\caption{(a) Integrated density of states $N$ (normalized to unity) vs.
energy. The full line corresponds to the discrete Cantor spectrum of
Fig.~1a with $r_1=0.34$, $r_2=0.20$, and $n$=12.  The dashed line is
given by $\epsilon^{d}$, with $d=-\ln 2/\ln r_1$, the spectral
dimension. (b) Spectral fluctuations. $N(\epsilon)$ after dividing by
$\epsilon^d$ (full line) and the smooth approximation (dotted line)
$\overline{N}/\epsilon^d= a + b\cos(\omega \ln \epsilon -\phi) $, where
$\omega=-2\pi/\ln r_1$. $a$ and $b$ are determined by requiring that
the exact and the smoothed fluctuations have the same average value and
variance. The condition of maximum overlap fixes $\phi$.}
\end{figure}
In the low temperature regime ($T \ll 1$), the expression above tends 
to the the scaling relation
\begin{equation} 
Z_{\cont}(\beta) =  \frac{r_1}{r_1+r_2} Z_{\cont}(\beta \, r_1)  ~. 
\label{eqZcont}   
\end{equation}
Note that, as in the discrete case, the scale factor $r_1$ is
responsible for the period of the log--oscillations. Thus the discrete
and continuous characteristic frequencies are equal.  The essencial
difference is the presence of $r_2$ in (\ref{eqZcont}), which can be
traced back to a different distribution of the spectral density (of
course, when $r_1=r_2$ both cases coincide).  In consequence, $r_2$
will also affect the mean value of specific heat, which is easily 
shown to be now
\begin{equation}
\langle C_{\cont} \rangle =  1-\ln(r_1+r_2)/\ln r_1 \equiv d' ~.
\label{Ccont}
\end{equation}
Equality (\ref{Ccont}) defines a new dimension $d'$, which, together
with $d$ and $d_f$, constitute the basic set of characteristic
dimensions of our problem.  We remark that these dimensions assume
different values, except for the particular case $r_1=r_2$.  Even
though the mean value (\ref{Ccont}) differs from its discrete
counterpart (\ref{meanheat}), the continuous and discrete specific
heats oscillate in a similar way about their respective averages.
However, the small--$r_1$ assymetries are more pronounced in the
continuous case. These facts can be appreciated by comparing Fig.~3a
and Fig.~3b.
  
The previous analysis for the two--scale spectrum (either discrete or
continuous) can also be generalized to the multiscale case. The
construction of a multiscale fractal spectrum starts from an arbitrary
discrete or continuous set of levels in the interval $[0,1]$ ($n=0$).
Then one makes $M$ rescaled copies of the pattern $n=0$, with different
scale factors $r_1,\ldots,r_M$. Each one of this copies is placed in
the unit interval at positions $a_1,\ldots,a_M$, so that the copies do
not overlap (this requires $a_i+r_i < a_{i+1}$). Iteration of this rule
eventually leads to a fractal of dimension $\sum_{i=1}^{M} r_i^{d_f}
=1$.  Analogous considerations to those made for the two--scale case
result in the following relationships for the discrete and the
continuous multiscale cases, respectively
\begin{eqnarray}
Z_{\disc}(\beta) & = & \frac{1}{M} 
                      \sum_{i=1}^M e^{-\beta a_i} 
                      Z_{\disc}(\beta r_i) ~,   \\    
Z_{\cont}(\beta) & = & \frac{1}{\sum_{i=1}^M  r_i} 
                      \sum_{i=1}^M r_i e^{-\beta a_i} 
                      Z_{\cont}(\beta r_i) ~.
\end{eqnarray}
These partition functions lead to the average specific heats 
($T \ll 1$)
\begin{eqnarray}
\langle C_{\disc} \rangle & = &   - \ln M / \ln r_1  \equiv d ~, \\
\langle C_{\cont} \rangle & = & 1 - \ln 
         \left( \sum_{i=1}^M r_i \right) / \ln r_1  \equiv d' ~.
\end{eqnarray}
(Naturally, the $d$ and $d'$ we introduced in Section II are the $M=2$
particular case of those defined above.) Once more the scaling
exponents only depend on $r_1$ in the discrete case and on the whole
set of scaling factors $r_j$ in the continuous (banded) case.  As in
the two--scale case, these scale factors will be the essential
ingredients for a very good approximate description of the
thermodynamics of the system.

We point out that the fractals considered in this paper might also be
analized in their {\em outbound} and {\em complete} versions (in the
nomenclature of \cite{tsallis}). These variations, which can also be
treated within our formalism, will give rise to a thermodynamics
analogous to that described above.

\section{Conclusions}

The models we have studied suggest that the hierarchical organization
of the energy spectra reflects itself in the specific heat in two
ways.  Simple scaling arguments show that the average behaviour is
associated to a non--integer spectral dimension ($d$ and $d'$ in our
examples), which in general is different from the fractal dimension
($d_f$). The corrections to this result are log--periodic oscillations
which can be traced back to the log--periodicity of the spectral
staircase. The number of oscillations that can be observed is related
to the hierarchical depth of the fractal spectrum, implying that these
anomalies may appear in systems displaying a self--similar spectrum up
to a {\em finite} hierarchichal depth.  These observations, related to
multi-scale fractal spectra, might also be relevant in the case of the
more realistic multifractal ones, because usually a few scales suffice
for a good description of a multifractal spectrum.
  
Moreover, since the effect is a consequence of the scale invariance of
the spectrum it is quite plausible that similar fenomena would
generically exist for bosonic and fermionic systems\cite{celia}.
 To support this conjecture, let us mention that   
Petri and Ruocco \cite{petri} have observed fractional scaling 
laws when studying the (Debye) vibrational specific heat of a
one--dimensional hierarchical model. However, those authors were mainly
concerned with mean values and did not discuss the small amplitude
oscillations that can be clearly observed in their results.  

Even though it is not surprising that log--periodic corrections (or
``complex exponents'') are a natural consequence of discrete scale
invariance \cite{sornette}, a contribution of the present paper is to
have reported and analyzed examples in which the connection between
scale invariance (of an energy spectrum) and log--periodicity (of the
specific heat as a function of temperature) shows up transparently.

Before concluding, let us comment on a possible connection of the
present calculation with the recently introduced nonextensive
thermostatistics\cite{tsallis2}.  Alemany\cite{alemany} has 
suggested that this formalism could be connected to systems with
fractally structured Boltzmann-Gibbs probability distributions.
Although, for our present calculation, we have not succeeded in making
a transparent connection along Alemany's lines, it is worthy mentioning
one intriguing feature. The generalized specific heat
$C_q(T)$ of the quantum one-dimensional harmonic 
oscillator\cite{pury} {\it does
present oscillations\/} if the entropic index $q$ satisfies $q<1$. 
In fact,
$C_q(T)/T^{1-q}$ is an oscillatory function of $T$, in a similar way
$C(T)$ is a periodic function of $\ln T$.

\acknowledgements
We gratefully acknowledge partial finantial support by CNPq,
FAPERJ and PRONEX (Brazilian Agencies). R. O. V. thanks
C. Anteneodo for useful discussions.

\end{document}